\documentclass[aps,prl,amsmath,amssymb,longbibliography,noeprint,10pt,twocolumn]{revtex4-2}
\usepackage{hyperref}
\usepackage{dutchcal}
\usepackage{color}
\usepackage{mathtools}
\usepackage{bbm}
\usepackage{enumitem}

\def\be{\begin{equation}}
\def\ee{\end{equation}}

\newcommand{\ch}{{\mathcal{h}}} 
\newcommand{\cC}{{\mathcal{C}}}

\newcommand{\muu}{\mu_{\rm u}}

\newcommand{\bH}{{\boldsymbol{H}}} 
\newcommand{\bC}{{\boldsymbol{C}}}

\newcommand{\bZ}{{\boldsymbol{Z}}}
\newcommand{\bmu}{{\boldsymbol{\mu}}}

\newcommand{\bmul}{\bmu_l}

\newcommand{\oH}{{\overline{H}}}
\newcommand{\oBH}{\boldsymbol{\overline{H}}}
\newcommand{\bP}{{\boldsymbol P}}
\newcommand{\npair}{N_{\rm pair}}
\newcommand{\npul}{N_{\rm pul}}

\newcommand{\bCi}{{\boldsymbol{C^{-1}}}} 

\newcommand{\bGi}{{\boldsymbol{G^{-1}}}}

\newcommand{\NC}{N_{\rm cr}}
\newcommand{\NF}{N_{\rm freq}}
\newcommand{\eq}[1]{(\ref{#1})}

\newcommand{\LP}{{\rm P}}
\newcommand{\Le}{L_{\rm eff}}
\newcommand{\bd}{\boldsymbol{d}}
\newcommand\scalemath[2]{\scalebox{#1}{\mbox{\ensuremath{\displaystyle #2}}}}

\newcommand{\bhc}{\boldsymbol{\hat c}}
\newcommand{\cG}{\mathcal{G}}
\newcommand{\cGi}{\mathcal{G}^{-1}}
\newcommand{\bcC}{\boldsymbol{\cC}}
\newcommand{\bcG}{\boldsymbol{\cG}}
\newcommand{\bcGi}{\boldsymbol{\cG^{-1}}}
\newcommand{\Fi}{F^{-1}}
\newcommand{\bD}{\boldsymbol{\Delta}}
\newcommand{\bF}{{\boldsymbol{F}}} 
\newcommand{\bFi}{{\boldsymbol{F^{-1}}}}
\newcommand{\bc}{\boldsymbol{c}}

\begin{document}

\title{Harmonic spectrum of pulsar timing array angular correlations}

\date{\today}

\author{Bruce Allen}
\email{bruce.allen@aei.mpg.de}
\affiliation{Max Planck Institute for Gravitational Physics (Albert
  Einstein Institute), Leibniz Universit\"at Hannover, Callinstrasse 38,
  D-30167, Hannover, Germany}

\author{Joseph D.\ Romano}
\email{joseph.romano@utrgv.edu}
\affiliation{Department of Physics and Astronomy,
University of Texas Rio Grande Valley,
One West University Boulevard,
Brownsville, TX 78520, USA}

\begin{abstract} \noindent Pulsar timing arrays (PTAs) detect gravitational waves (GWs) via the correlations they create in the arrival times of pulses from different pulsars.  The mean correlation, a function of the angle $\gamma$ between the directions to two pulsars, was predicted in 1983 by Hellings and Downs (HD).  Observation of this angular pattern is crucial evidence that GWs are present, so PTAs ``reconstruct the HD curve'' by estimating the correlation using pulsar pairs separated by similar angles. The angular pattern may be also expressed as a ``harmonic sum'' of Legendre polynomials $\LP_l(\cos \gamma)$, with coefficients $c_l$.  Here, assuming that the GWs and pulsar noise are described by a Gaussian ensemble, we derive optimal estimators for the $c_l$ and compute their variance.  We consider two choices for ``optimal''.  The first minimizes the variance of each $c_l$, independent of the values of the others. The second finds the set of $c_l$ which minimizes the (squared) deviation of the reconstructed correlation curve from its mean.  These are analogous to the so-called ``dirty'' and ``clean''  maps of the electromagnetic and (audio-band) GW backgrounds.
\end{abstract}


\maketitle

{\it Introduction.---} This paper is about reconstruction of the
Hellings and Downs (HD) correlation~\cite{HD}, starting from pulsar
timing array (PTA)
data~\cite{EPTA_new23,NANOGrav_new,PPTA_new,CPTA_new,MeerKAT}.  Previous
work~\cite{OptimalHD} shows how to optimally estimate this correlation
using data from a set of pulsar pairs lying in a particular angular
separation
bin~\cite{AllenRomanoPTA}.  This paper shows how to optimally
reconstruct the correlation expressed in harmonic form, as a sum of
Legendre polynomials, using all pulsar pairs.

This paper must be read in conjunction with~\cite{OptimalHD}, because
we omit almost all equations which can be taken from there.  Those are
referred to using the notation (AR7) to indicate Eq.~(7) of
~\cite{OptimalHD}.

The goal is to reconstruct the HD correlation $-1 \le
\frac{3}{2} \mu \le 1$ as a function of the angle $\gamma \in [0,\pi]$
between the directions to pulsars~\cite[Introduction]{OptimalHD}. The
correlation can be expressed as a sum of Legendre polynomials
\be          
\mu(\gamma)
= \sum_{l=0}^\infty c_l \LP_l(\cos\gamma) \, .
\label{e:HD_harmonic}
\ee
Here, we derive optimal estimators for the $c_l$ and compute their
variance.

\begin{figure}
\centering
\includegraphics[width=0.45\textwidth]{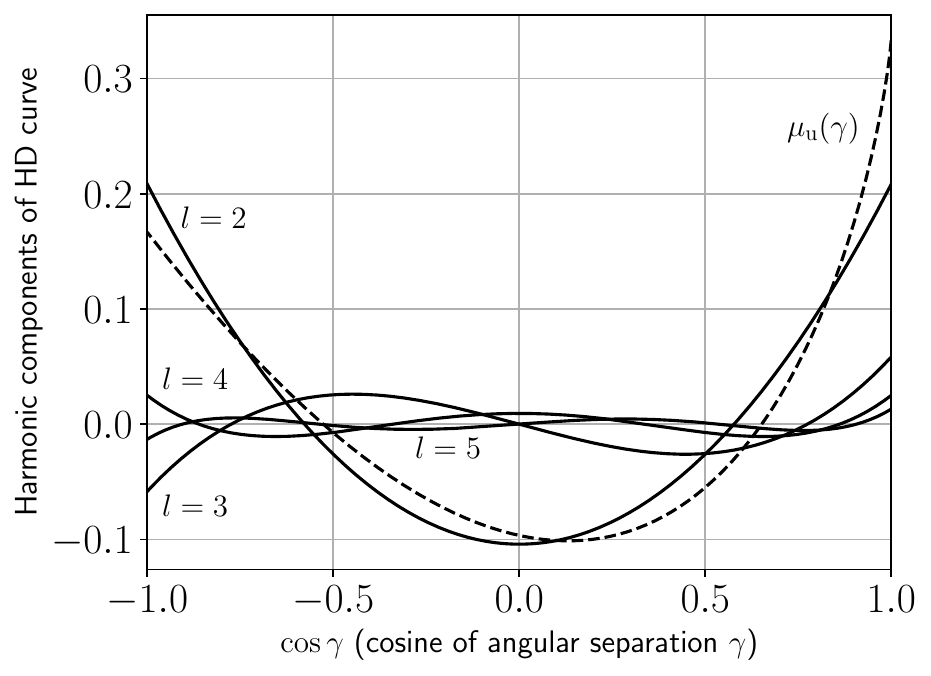}
\caption{The HD curve $\muu(\gamma)$ (dashed line) and its first four
  harmonic components $\langle c_l \rangle \LP_l(\cos \gamma)$ (solid
  lines).}
  \label{f:plots}
\end{figure}

The expected value of the correlation is called the HD curve
$\muu(\gamma) \equiv \langle \mu(\gamma) \rangle$ and is illustrated
in Fig.~\ref{f:plots}.  The corresponding expected values of the
harmonic coefficients are computed
in~\cite{gair, Roebber_2017,allen2024harmonic} and are
\begin{equation}
\langle c_l \rangle =  \begin{cases}
    0 & \text{ for } l < 2 \\
    (2l+1)/\bigl( (l+2)(l+1)l(l-1) \bigr) & \text{ for } l \ge 2 \, .
    \label{e:a_l}
  \end{cases}
\end{equation}
These different contributions are also shown in Fig.~\ref{f:plots};
partial sums may be seen in~\cite[Fig.~11]{FAQ}. Because the $\langle
c_l \rangle$ fall off like $1/l^3$ for large $l$, summing the terms up
to $l=5$ already gives a good approximation of the HD curve.  (Note
that the correlation is doubled to $3\muu(0) = 1$ for pulsars that are
closer together than the typical GW
wavelength, see~\cite[App.~C.2]{AllenHDVariance}.)

{\it Approach 1: Matched Filter --} Our first approach is inspired by
that of~\cite{OptimalHD}.  In (AR23) we considered estimators $\hat
\mu(\gamma)$ which are arbitrary linear combinations of the form
\be  
  \label{e:estimator}
   \sum_{ab \in \gamma} \sum_{j,k} W_{ab}^{jk} Z_a^j Z_b^k \, ,
\ee
where $Z_a^j$ is the (complex) Fourier amplitude of the redshift of
pulsar $a$ in frequency bin $j$, defined in~(AR12), and the $
W_{ab}^{jk}$ are a set of weights.  There, the sum was over a set of
pulsar pairs $ab$ in a narrow angular bin at angle $\gamma$, and the
weights satisfied $W_{ab}^{jk*} = W_{ab}^{-j,-k}$, to ensures that the
estimator is real.  The weights were selected so that the estimator is
unbiased and minimum variance.  (Note that all analysis there and here
may also be done in terms of timing residuals; the results are
identical, see~\cite[Footnote 19]{OptimalHD}.)

In analogy with this, we define an unbiased estimator for $c_l$. The
estimator is analogous to (AR28), although we do not know how to
derive it from first principles.  For a PTA containing $\npul$
pulsars, the estimator is
\begin{eqnarray}
  \label{e:cl1}
  \hat c_l  & \equiv & \langle c_l \rangle
  \frac{(\bmu_l \oBH)^t \bCi \boldsymbol{Z Z} }{(\bmu_l \oBH)^t \bCi (\bmu \oBH) } \\[7pt]
  \nonumber
  & \equiv & \langle c_l \rangle
  \frac{\displaystyle \sum_{a<b} \sum_{c<d} \sum_{j,k} \sum_{\ell, m}
    \mu_{l,ab} \oH_{jk} \bigl( C^{-1} \bigr)_{ab,cd}^{jk,\ell m} Z_c^\ell Z_d^m  \,\,\,\,}
{\displaystyle \sum_{a<b} \sum_{c<d} \sum_{j,k} \sum_{\ell, m}
  \mu_{l,ab} \oH_{jk} \bigl( C^{-1} \bigr)_{ab,cd}^{jk,\ell m} \mu_{cd} \oH_{\ell m} }\, ,
\end{eqnarray}
where $\boldsymbol{Z Z} \equiv Z_a^j Z_b^k$.  Here,
\be
  \label{e:mu_l}
\bmul \equiv \mu_{{}_{l,ab}} \equiv (1 + \delta_{ab}) \LP_l(\cos \gamma_{ab}) \,.
\ee
and $\bmu = \sum_l \langle c_l \rangle \bmul$ are vectors of dimension
$\npair = \npul (\npul - 1)/2$.  In contrast with \eq{e:estimator},
the sum is \emph{not} confined to a particular angular bin, but
includes \emph{all possible} pulsar-pair cross-correlations $a<b$.

This estimator takes the form
\be
\hat c_l = \langle c_l\rangle\frac{d_l}{u_l}\,,
\label{e:cl2}
\ee
where we have defined
\begin{align}
d_l &\equiv (\bmu_l \oBH) ^t \bCi \boldsymbol{ZZ} \, ,
\label{e:dl}
\\
u_l &\equiv (\bmu_l \oBH) ^t \bCi (\bmu\oBH) \, .
\label{e:ul}
\end{align}
Expression \eq{e:dl} is an \emph{optimal matched filter} for the
angular correlation pattern $\bmu_l$. The corresponding ``filter
template'' \eq{e:mu_l} is proportional to the $l$'th term in
\eq{e:HD_harmonic}; the $\delta_{ab}$ can be dropped since $a<b$.
This is an optimal matched filter because it is obtained by
``whitening the template and the data'' via the inverse covariance
matrix $\bCi$, and then normalizing.  Our normalization, placing $u_l$
in the denominator of \eq{e:cl2}, gives an unbiased estimator of
$c_l$.  To see this, note that $\langle \boldsymbol{Z Z} \rangle =
\bmu \oBH$. It follows from inspection of \eq{e:dl} and \eq{e:ul} that
$\langle d_l\rangle = u_l$, thus implying $\langle \hat c_l \rangle =
\langle c_l\rangle$.

The variance of the optimal matched filter estimator
  \eq{e:cl1} is easy to calculate.  The covariance of $\boldsymbol{Z
    Z}$ is 
\begin{equation}
  \bcC  \equiv \bigl\langle \boldsymbol{Z Z} (\boldsymbol{Z Z})^\dagger \bigr\rangle
  -
  \bigl\langle \boldsymbol{Z Z} \bigr\rangle \bigl\langle (\boldsymbol{Z Z})^\dagger \bigr\rangle  \,.
\label{e:bC}
\end{equation}
An explicit formula for 
$\bcC\equiv \cC_{ab,cd}^{jk,\ell m}$ 
is given in (AR19), in terms of covariance matrices computed from the
gravitational wave background (GWB) and pulsar noise power spectra. 
The covariance matrix $\bC$, whose inverse $\bCi$ appears in \eq{e:cl1},
is the symmetric part of $\bcC$ with respect to the frequency 
indices (AR20):
\be
\bC\equiv C_{ab,cd}^{jk,\ell m}
\equiv \cC_{ab,cd}^{(jk),(\ell m)}
\ee
From the definition \eq{e:dl} of the $d_l$ 
it follows that their covariance is
\begin{eqnarray}
  \nonumber
&&\langle d_{l} d_{l'} \rangle -  \langle d_{l} \rangle \langle d_{l'} \rangle
\\   \nonumber
&&\hspace{0.2in} = (\bmu_l\oBH)^t \bCi 
\left(\langle (\bZ\bZ)(\bZ\bZ)^\dagger\rangle - \langle \bZ\bZ\rangle\langle(\bZ\bZ)^\dagger\rangle\right) 
\\  \nonumber
&&\hspace{1.2in} \times \bCi (\bmu_{l'}\oBH)
\\  \nonumber
&&\hspace{0.2in}
= (\bmu_l\oBH)^t \bCi \bC \bCi (\bmu_{l'}\oBH)
\\
&&\hspace{0.2in} = F_{ll'}\,,
\label{e:covd}
\end{eqnarray}
where \eq{e:bC} was used to get the second equality and we have
defined a matrix
\be
\bF \equiv F_{ll'} \equiv (\bmu_l \oBH) ^t \bCi (\bmu_{l'} \oBH) \,,
\label{e:Fll'}
\ee
where $l$ and $l'$ are indices labeling harmonics.  Thus, the variance
of the HD correlation estimator is
\begin{equation}
\begin{aligned}
\label{e:var_c1}
 \sigma^2_{\hat c_l} 
&\equiv \langle \hat c^2_l \rangle - \langle \hat c_l \rangle^2 
\\
&= \langle c_l\rangle^2 \bigl(\langle d_l^2\rangle - \langle d_l\rangle^2 \bigr)/u_l^2
\\
&= \langle c_l\rangle^2 F_{ll}/u_l^2\,,
\end{aligned}
\end{equation}
where $F_{ll}$ are the diagonal elements of $\bF$.


We will write the variance \eq{e:var_c1} as
\be
\sigma^2_{\hat c_l} 
=\frac{\langle c_l\rangle^2}{(2\Le + 1)\NF}
= \frac{\langle c_l\rangle^2 F_{ll}}{u_l^2}\,,
\label{e:var_c2}
\ee
where $\NF$ and $2\Le +1$ are functions of $l$.  (This, along with the
formulas given below for $\NF$ and $2\Le +1$, is one of the two main
results of this paper.)  The division into the two factors is
arbitrary; we break this degeneracy by imposing two requirements: (i)
The effective number of frequency bins $\NF$ for which the GWB
dominates pulsar noise equals $\NC$ in the crossover frequency
limit. (ii) The effective number of harmonic degrees of freedom $2\Le
+1$ depends only on the ``geometry'' of the pulsar timing array,
meaning the pulsar sky directions.  Note: the crossover frequency
limit is discussed and defined in~\cite{OptimalHD} between (AR34) and
(AR35).

To find $2\Le+1$ and $\NF$, we start with requirement (i), computing
$\bF$ in the crossover frequency limit, where below ``crossover''
frequency $\NC/T$, the GWB dominates the redshifts or timing
residuals, and above that frequency, the pulsar noise dominates.  In
this crossover limit, $\NF \to \NC$.  The spectrum $H_{jk}$ of the GWB
is nonzero only for frequency bins $|j|,|k| \le \NC$ and vanishes if
either $|j|$ or $|k|$ is larger than $\NC$.  The pulsar spectrum
$P^{jk}_a$ has the opposite behavior.  For this case, the inverse of
the covariance matrix is given by (AR35).  Since $\bH$ or $\oBH$
contracted with $\bP_a$ vanishes, the same calculation as in (AR37)
shows that
\be
F_{ll'} \to 2\NF\,\cG_{ll'}\,,
\qquad
u_l \to 2\NF\,v_l\,,
\label{e:cross1}
\ee
where
\begin{align}
&\bcG \equiv \cG_{ll'} \equiv (\bmu_l)^t \bGi \bmu_{l'} \,,
\label{e:cGll}
\\
&v_l \equiv (\bmu_l)^t \bGi \bmu\,.
\label{e:vl}
\end{align}
Substituting these
limiting expressions for $F_{ll}$ and $u_l$ into the rhs of
\eq{e:var_c2} implies that
\be
2\Le + 1 \equiv \frac{2 v_l^2}{\cG_{ll}}\,.
\label{e:Le1}
\ee
While \eq{e:Le1} was obtained in the crossover limit, 
requirement (ii) implies that it must always hold.  
It then follows from  \eq{e:var_c2} and \eq{e:Le1} that
\be
\NF \equiv \frac{1}{2}\frac{\cG_{ll}}{F_{ll}}\left(\frac{u_l}{v_l}\right)^2
\label{e:NF1}
\ee
holds in general (not just in the crossover limit).  Note that $\NF$
depends upon the GWB and pulsar power spectra, as well as upon the
pulsar sky positions and $l$.

The effective number of harmonic degrees of freedom $2\Le +1$ given by 
\eq{e:Le1} simplifies when there are many pulsars uniformly distributed on the sky. 
To see this, we use results from
\cite{AllenRomanoPTA}, noting that the $C_l$ as defined in
\cite[Eq.~(4.14)]{AllenRomanoPTA} are related to the $\langle c_l
\rangle$ as given in \eq{e:a_l} by $\langle c_l \rangle = (2l+1) C_l$.
From \cite[set $\ch=1$ in Eq.~(4.25)]{AllenRomanoPTA} the many-pulsar-pair limit of
$\bGi \equiv (G^{-1})_{ab,cd}$ is proportional to
\begin{equation}
  \begin{aligned}
  \label{e:cl5}
  G^{-1}(x,x')
  & = \frac{1}{8} \sum_{l=2}^\infty \frac{(2l+1)^3}{\langle  c_l \rangle ^2} \LP_l(x) \LP_l(x') \, ,
  \end{aligned}
\end{equation}
where $x = \cos \gamma_{ab}$ and $x'=\cos \gamma_{cd}$, and it is understood
that this denotes the average value over many pulsar pairs at a given
angular separation.
The discrete and continuous quantities
and their inverses are related by~\cite[Eq.~(4.29)]{AllenRomanoPTA} 
\be
\begin{aligned}
G_{ab,cd} &= G(\cos\gamma_{ab},\cos\gamma_{cd})
\\
(G^{-1})_{ab,cd} &= \left(\frac{2}{\npair}\right)^2 
G^{-1}(\cos\gamma_{ab}, \cos\gamma_{cd})\,.
\end{aligned}
\ee
In the many-pulsar limit, $\frac{1}{\npair}\sum_{a<b} \to
\frac{1}{2}\int_{-1}^1 dx$.  So, by using \eq{e:mu_l} and \eq{e:cGll}
one can see that $\cG_{ll'}$ approaches
\begin{equation}
  \label{e:cl6}
  \begin{aligned}
    \cG_{ll'} & \to \int_{-1}^1 \!\!\!\!\! dx \! \int_{-1}^1  \! \!\!\!\!\! dx' \,
    \LP_l(x) \, G^{-1}(x,x') \, \LP_{l'}(x')  \\
    & = \begin{cases}
      0 & \text{ if } l<2 \\[2mm]
      \displaystyle \frac{2l+1} {2 \langle c_l \rangle^2}\delta_{ll'} & \text{ if } l\ge 2 \, .
      \end{cases}
\end{aligned}
\end{equation}
To obtain the final equality above we have used \eq{e:cl5} and the
orthogonality relation
\begin{equation}
  \label{e:cl7}
   \int_{-1}^1 \!\!\!\!\! dx \, \LP_l(x) \LP_{l'}(x) = \frac{2}{2l+1}
   \delta_{ll'}
\end{equation}
for the Legendre polynomials.

The diagonal form of $\bcG$ in this many pulsar limit reflects the
orthogonality of the different spherical harmonics when sampled very
finely on the sphere.  Note that in this limit $\bcG$ vanishes along
the first two rows and columns $l=0,1$ and $l'=0,1$. This is because
$\bGi$ does not contain either of these harmonics, as can be seen from
\eq{e:cl5}.  It follows that in this limit, every entry of the
pseudoinverse $\bcGi$ of $\bcG$ vanishes except for the diagonal
elements starting in the third row/column, where
$(\cGi)_{ll}=1/\cG_{ll}$ for $l\ge 2$.

We use a similar calculation to evaluate $v_l$ in the many-pulsar limit.
Starting from \eq{e:vl} gives
\begin{equation}
  \label{e:cl9}
  v_l 
  \to \cG_{ll} \langle c_l \rangle \to \frac{2l+1}{2\langle c_l\rangle}\ \ \text{for }
l\ge 2\,.
\end{equation}
Inserting \eq{e:cl6} and \eq{e:cl9} into \eq{e:Le1} implies
that in the limit of many pulsars, uniformly distributed on the sky,
\begin{equation}
  2 \Le + 1  \to
  \begin{cases}
    0 & \text{ for } l <2 \\
    2 l  + 1 & \text{ for } l \ge 2 \, .
  \end{cases}
\end{equation}
In effect, with enough pulsars spread around the sky, all of the
degrees of freedom of any multipole can be observed and measured.  If
not, then \eq{e:Le1} is the effective number of degrees of freedom
which could be observed with the given set, if enough SNR were
available.  Shown in Table~\ref{t:NG15Lvalues} are the values of $ 2
\Le + 1$ for the pulsars used by the different pulsar timing arrays
mentioned in the Introduction, along with a fictitious PTAs formed from
the union of their pulsars, and larger numbers of randomly-placed
pulsars.

\begin{table*}
  \begin{tabular}{cccccccccc}
 &  & EPTA 25 & PPTA 30 & CPTA 57 & NG 67 & IPTA3 115 & All 127 & RU 200 & RU 300 \\
$l$ & $2l+1$ & $\hspace{1.2em} 2L_e + 1 \hspace{1.2em} $ & $\hspace{1.2em} 2L_e + 1 \hspace{1.2em} $ & $\hspace{1.2em} 2L_e + 1 \hspace{1.2em} $ & $\hspace{1.2em} 2L_e + 1 \hspace{1.2em} $ & $\hspace{1.2em} 2L_e + 1 \hspace{1.2em} $ & $\hspace{1.2em} 2L_e + 1 \hspace{1.2em} $ & $\hspace{1.2em} 2L_e + 1 \hspace{1.2em} $ & $\hspace{1.2em} 2L_e + 1 \hspace{1.2em} $ \\
\hline
0  &  1 & $0.067\,\,\phantom{0.}0\phantom{00}$ & $0.001\,\,\phantom{0.}0\phantom{00}$ & $0.030\,\,\phantom{0.}0\phantom{00}$ & $0.009\,\,\phantom{0.}0\phantom{00}$ & $0.002\,\,\phantom{0.}0\phantom{00}$ & $0.001\,\,\phantom{0.}0\phantom{00}$ & $0.001\,\,\phantom{0.}0\phantom{00}$ & $\phantom{0.}0\phantom{00}\,\,\phantom{0.}0\phantom{00}$ \\
1  &  3 & $0.011\,\,\phantom{0.}0\phantom{00}$ & $0.057\,\,\phantom{0.}0\phantom{00}$ & $0.267\,\,\phantom{0.}0\phantom{00}$ & $0.022\,\,\phantom{0.}0\phantom{00}$ & $0.010\,\,\phantom{0.}0\phantom{00}$ & $0.008\,\,\phantom{0.}0\phantom{00}$ & $0.001\,\,\phantom{0.}0\phantom{00}$ & $\phantom{0.}0\phantom{00}\,\,\phantom{0.}0\phantom{00}$ \\
2  &  5 & $2.015\,\,1.725$ & $2.628\,\,2.310$ & $3.246\,\,2.889$ & $3.388\,\,3.271$ & $4.023\,\,3.921$ & $4.122\,\,4.069$ & $4.551\,\,4.560$ & $4.709\,\,4.710$ \\
3  &  7 & $0.604\,\,0.362$ & $0.932\,\,0.450$ & $1.763\,\,1.037$ & $1.908\,\,1.669$ & $2.912\,\,2.615$ & $3.189\,\,3.004$ & $4.551\,\,4.586$ & $5.265\,\,5.270$ \\
4  &  9 & $0.386\,\,0.064$ & $0.363\,\,0.107$ & $1.035\,\,0.304$ & $0.788\,\,0.499$ & $1.514\,\,1.114$ & $1.641\,\,1.376$ & $2.999\,\,3.071$ & $4.207\,\,4.221$ \\
5  & 11 & $0.250\,\,0.017$ & $0.081\,\,0.030$ & $0.557\,\,0.120$ & $0.464\,\,0.183$ & $0.914\,\,0.460$ & $0.906\,\,0.575$ & $1.474\,\,1.537$ & $2.523\,\,2.549$ \\
6  & 13 & $0.097\,\,0.006$ & $0.006\,\,0.007$ & $0.306\,\,0.048$ & $0.271\,\,0.072$ & $0.536\,\,0.191$ & $0.525\,\,0.242$ & $0.640\,\,0.692$ & $1.270\,\,1.295$ \\
7  & 15 & $0.108\,\,0.002$ & $0.002\,\,0.003$ & $0.179\,\,0.020$ & $0.141\,\,0.031$ & $0.307\,\,0.086$ & $0.299\,\,0.108$ & $0.270\,\,0.309$ & $0.601\,\,0.622$ \\
8  & 17 & $0.019\,\,0.001$ & $0.008\,\,0.001$ & $0.111\,\,0.009$ & $0.084\,\,0.014$ & $0.196\,\,0.039$ & $0.197\,\,0.049$ & $0.118\,\,0.142$ & $0.288\,\,0.298$ \\
9  & 19 & $0.031\,\,0.000$ & $0.016\,\,0.001$ & $0.068\,\,0.004$ & $0.038\,\,0.007$ & $0.104\,\,0.020$ & $0.102\,\,0.025$ & $0.053\,\,0.068$ & $0.139\,\,0.151$ \\
10 & 21 & $0.018\,\,0.000$ & $0.026\,\,0.000$ & $0.044\,\,0.002$ & $0.018\,\,0.003$ & $0.059\,\,0.010$ & $0.058\,\,0.013$ & $0.022\,\,0.034$ & $0.073\,\,0.078$ \\
11 & 23 & $0.009\,\,0.000$ & $0.034\,\,0.000$ & $0.026\,\,0.001$ & $0.004\,\,0.002$ & $0.028\,\,0.006$ & $0.031\,\,0.007$ & $0.008\,\,0.018$ & $0.040\,\,0.043$ \\
12 & 25 & $0.005\,\,0.000$ & $0.020\,\,0.000$ & $0.028\,\,0.001$ & $0.003\,\,0.001$ & $0.017\,\,0.003$ & $0.023\,\,0.004$ & $0.003\,\,0.011$ & $0.023\,\,0.024$ 
  \end{tabular}
  \caption{\label{t:NG15Lvalues} The effective number of observable
    angular degrees of freedom $2 \Le + 1$ for different PTAs, which
    are listed on the first row followed by the number of pulsars.  In
    each entry, the first value is \eq{e:Le1} for Approach 1, and the
    second value is \eq{e:Le2} for Approach 2. (By definition, the
    latter vanish for $l<2$.) The main sensitivity is to the first few
    harmonics beginning with $l=2$. The ``All'' PTA includes all
    pulsars from the previous columns, eliminating duplicates. The
    final two columns show 200 and 300 pulsars placed randomly on the
    sky as a Random Uniform Poisson process.}
\end{table*}

Thus, in the many pulsar case, we obtain
\begin{equation}
  \label{e:cl11}
  \sigma^2_{\hat c_l} = \frac{\langle c_l \rangle^2}{(2l +1) \NF} \, .
\end{equation}
This is a generalization of the many-pulsar single-frequency-bin case
considered by Roebber and Holder~\cite{Roebber_2017}, which is
obtained by setting $\NF=1$.

{\it Approach 2: Best Fit --} Our second approach differs from the
first.  There, the goal was to estimate a single $c_l$ ``in
isolation'', without taking into consideration the values of the other
$c_{l'}$ (for $l' \ne l$). Here, we will instead minimize a ``global''
quantity $\chi^2$ that characterizes the difference between the
observed correlations and the harmonic model \eq{e:HD_harmonic}.  This
strives to simultaneously find the best estimator for a \emph{set} of
$c_l$.  We denote this set by $\bc \equiv \{ c_l \}$, with the understanding
that this denotes some finite subset of $\{c_0, c_1, \dots\, \}$.  (The remaining $c_l$ may be set to
zero or other fixed/definite values.)

The starting point of this $\bc$ ``reconstruction'' is a real
quadratic function $\chi^2$ of $\bc$, defined by
\begin{eqnarray}
\label{e:n1}
\chi^2(\bc) & \equiv &  \bD^t(\bc) \bCi \bD^* (\bc) \\
\nonumber
         & \equiv &\sum_{a<b}  \sum_{c<d} \sum_{j,k}  \sum_{\ell,m}
  \Delta_{ab}^{jk}(\bc)\bigl(C^{-1}\bigr)_{ab,cd}^{jk,\ell m}
  \Delta_{cd}^{\ell m}{}^*(\bc)\,,
\end{eqnarray}
where 
\be
\label{e:n1a}
\bD (\bc) \equiv
\Delta_{ab}^{jk}(\bc) \equiv Z_a^j Z_b^k -  \oH_{jk} \scalemath{0.85}{\sum_{c_l \in \bc}} c_l \mu_{l,ab} \,
\ee
and $\bCi$ is the inverse of the covariance matrix given in \eq{e:bC}.
The $\Delta$'s and the statistic $\chi^2$ are measures of the difference
between the observed spatial correlations and the model $\sum_l c_l
\mu_{l,ab}$ of \eq{e:HD_harmonic}, restricted to $c_l \in \bc$.
The value of $\chi^2$ is a
function of the observed redshifts (or timing residuals), of the GWB
and pulsar noise spectral models, and of a set of real coefficients
$\bc$.

In this approach, model selection or model fitting is carried out by
finding the $\bc$ that minimize $\chi^2$, with the other quantities
held fixed.  The values of $\bc$ that minimize $\chi^2$ will be
denoted by $\hat \bc \equiv \{ \hat c_l \}$; they are estimators of the harmonic amplitudes
$\bc$ of the spatial correlations, which describe our particular
realization of the Universe, as defined by \eq{e:HD_harmonic}.  In
contrast with Approach 1, which minimizes the variance of $\hat c_l$
between different realizations of the Universe drawn from the Gaussian
ensemble, this approach minimizes the overall mismatch $\chi^2$
between the observed correlations in our Universe and the harmonic
decomposition \eq{e:HD_harmonic}. (The Gaussian ensemble is employed
to characterize how much mismatch will remain, on the average.)

To accomplish this minimization, take the partial derivative of
\eq{e:n1} with respect to a particular $c_{l} \in \bc$, and set it to zero:
\begin{equation}
  \frac{\partial \chi^2(\bc)}{\partial c_{l}}\biggr\rvert_{\bc = \hat \bc} = 0 \,.
\end{equation}
For any given value of $l$, this results in a set of linear equations
for the estimators
\begin{equation}
  \label{e:Fc=d}
  \sum_{\hat c_{l'} \in \hat \bc} F_{ll'}\hat c_{l'}  = d_l\,,
\end{equation}
where $F_{ll'}$ is defined in \eq{e:Fll'},
and $l$ ranges over the set of subscripts that appear in $\bc$.
To obtain \eq{e:Fc=d}, we combined terms, 
exploiting the fact that $\bCi$ is a symmetric matrix.

Note that the $d_l$, which were defined earlier in \eq{e:dl}, are real,
since $(\bmu_l\oBH)^t \bCi$ annihilates the imaginary part of 
$\bZ\bZ$.
This is a consequence of $\bmu_l$, $\oH$, and $\bCi$ all being real, 
which implies that 
\be
\oH_{jk}\bigl(C^{-1}\bigr)_{ab,cd}^{jk,\ell m}
=\oH_{-j,-k}\bigl(C^{-1}\bigr)_{ab,cd}^{-j,-k,-\ell,- m}\,.
\label{e:HCi_real}
\ee
Thus,
\begin{eqnarray}
  \nonumber
&&\!\!\!\! \sum_{jk}\sum_{\ell m} \oH_{jk}\bigl(C^{-1}\bigr)_{ab,cd}^{jk,\ell,m} Z_c^\ell Z_d^m
\\   \nonumber
&&\hspace{0.1in}
=\sum_{jk}\sum_{\ell m} \oH_{-j,-k}\bigl(C^{-1}\bigr)_{ab,cd}^{-j,-k,-\ell,-m} Z_c^{-\ell} Z_d^{-m}
\\   \nonumber
&&\hspace{0.1in}
=\sum_{jk}\sum_{\ell m} \oH_{jk}\bigl(C^{-1}\bigr)_{ab,cd}^{jk,\ell m} Z_c^{-\ell} Z_d^{-m}
\\
&&\hspace{0.1in}
=\sum_{jk}\sum_{\ell m} \oH_{jk}\bigl(C^{-1}\bigr)_{ab,cd}^{jk,\ell m}\Re(Z_c^{\ell} Z_d^{m})\,,
\end{eqnarray}
where the first equality follows from changing the sign of the dummy 
summation variables $j$, $k$, $\ell$, $m$; the second equality follows from 
\eq{e:HCi_real}; and the last equality follows from the previous two lines and 
the definition of the real part of $\boldsymbol{ZZ}$.

The number of linear equations \eq{e:Fc=d} is the same as
the number of $c_l$ in the finite set $\bc$.  
 We can choose which ones we wish to
estimate, and are free to arbitrarily set some (e.g., $\hat c_0$ and/or
$\hat c_1$) to zero (or other values) if desired.
In any case, to arrive at a unique
solution, the number of $\hat c_l$ to be determined must equal the
number of linearly independent equations.  Let us use $N_L$ to
denotes the number of $\hat c_l$ that we wish to obtain, which is the order
of the set $\bc$.
For example, if
by fiat we constrain $\hat c_0 = \hat c_1 = 0$, then we would obtain values for
$\hat c_2, \hat c_3, \cdots, \hat c_{{}_{N_L+1}}$.
The solution for the $\hat c_l$ are then given by 
\be
  \label{e:soln1}
  \bhc = \bFi \bd \, ,
\ee
where $\bFi$ is the matrix inverse (or pseudoinverse) of $\bF$.  Note
that from here forward, we treat $\bc$ as a column vector of dimension
$N_L$ rather than as a set.

The application of $\bFi$ to $\bd$ can be thought of as 
``deconvolving" the response of the measurement process 
in order to extract the harmonic components $c_l$.
In the context of map making, $\bF\equiv F_{ll'}$ is 
called the ``Fisher matrix" and the collection $\bd\equiv d_l$
is called the ``dirty map", see~\cite[Sec.~III.B]{Thrane_2009}.
The ``clean map" $\bhc \equiv \hat c_l$ is the end result of 
the deconvolution process.
Indeed, it is easy to see that the ``clean map" estimators $\bhc$ are unbiased. 
This follows from 
$\langle d_l\rangle= u_l$, which we discussed in the sentences 
following \eq{e:mu_l}, 
together with $u_l = \sum_{l'}F_{ll'}\langle c_{l'}\rangle$, which immediately
follows from the definitions \eq{e:ul} and \eq{e:Fll'} 
of $u_l$ and $F_{ll'}$.

The covariance of the estimators $\hat c_l$ is easily calculated
by first expressing it in terms of the covariance of the 
``dirty map" data $d_l$:
\be
\begin{aligned}
\sigma^2_{ll'} 
& \equiv \langle \hat c_l \hat c_{l'} \rangle -  \langle \hat c_l \rangle \langle \hat c_{l'} \rangle 
\\
& = \sum_{l_1}\sum_{l_2}
(\Fi)_{ll_1} 
\left(\langle d_{l_1} d_{l_2} \rangle -  \langle d_{l_1} \rangle \langle d_{l_2} \rangle\right)
(\Fi)_{l_2 l'} \\
& = \sum_{l_1}\sum_{l_2} (\Fi)_{ll_1} F_{l_1l_2} (\Fi)_{l_2 l'} \\
& = (\Fi)_{ll'}\,,
\label{e:covc2}
\end{aligned}
\ee
where we used \eq{e:covd} to obtain the third equality.
Note that if $F_{ll'}$ is not invertible, 
then $(\Fi)_{ll'}$ denotes its pseudoinverse, and the
product $\bFi \bF$ is a rank-$\bF$ projector onto the nonnull space.

The variance of a particular $\hat c_l$ estimator is obtained
by setting $l=l'$ in \eq{e:covc2}:
\begin{equation}
 \sigma^2_{\hat c_l} \equiv \sigma^2_{ll} =  (\Fi)_{ll}\,.
\label{e:var_c3}
\end{equation} 
In analogy with Approach 1, we write this variance
in the form
\be
\sigma^2_{\hat c_l} 
=\frac{\langle c_l\rangle^2}{(2\Le + 1)\NF}=(\Fi)_{ll}\,,
\label{e:var_c4}
\ee
where $\NF$ and $2\Le +1$ are functions of $l$.  
As noted for Approach 1, the division into the
two factors is arbitrary.
So we break this degeneracy by imposing the 
same two requirements as before.
(See the discussion after \eq{e:var_c2}.)

For requirement (i), we again consider the 
``crossover" frequency limit, where $\NF\to\NC$.
As before, we find $\bF \to 2\NF\bcG$.
But this time, we write this relation in terms 
of the diagonal components of the inverse 
matrices
\be
(\Fi)_{ll} \to \frac{(\cGi)_{ll}}{2\NF} \text{\ \ for } 
l\ge 2\,,
\label{e:cross2}
\ee
where $\bcGi$ is the pseudoinverse of $\bcG$.  
Substituting this limiting expression for $(\Fi)_{ll}$ 
into the rhs of \eq{e:var_c4} implies that
\be
2\Le + 1 \equiv \frac{2\langle c_l\rangle^2}{\left(\cGi\right)_{ll}}\,.
\label{e:Le2} 
\ee
As we argued for Approach 1, this relation must always 
hold as a consequence of requirement (ii).
It then follows from \eq{e:var_c4} that
\be
\NF \equiv \frac{1}{2}\frac{\left(\cGi\right)_{ll}}{\left(\Fi\right)_{ll}}\,.
\label{e:NF2}
\ee

As we saw for Approach 1, the effective number of harmonic degrees 
of freedom $2\Le +1$ simplifies when there are many pulsars uniformly 
distributed on the sky. In this limit \eq{e:Le2} gives
\be
2\Le+1\equiv\frac{2\langle c_l\rangle^2}{\left(\cGi\right)_{ll}}
\to 
2\langle c_l\rangle^2 \cG_{ll} \to 2l+1
 \text{\ \ for } 
l\ge 2\,,
\label{e:Le}
\end{equation}
as a consequence of \eq{e:cl6}.
This is precisely the number of degrees of freedom associated with harmonic index $l$.

Table~\ref{t:NG15Lvalues} shows values of $2\Le +1$ computed using
\eq{e:Le2} (with $N_L=16$) for several real and fictional PTAs.  We note
that when the number of pulsars is large, the values of $2\Le +1$ are
similar for both approaches.  We expect that $2\Le +1$ should satisfy
$2\Le +1 \le 2l+1$, but have not been able to prove this.  We would
have expected that adding pulsars to an array always increases $2\Le
+1$, but we have been unable to prove this and note that numerical
experiments indicate that it is not the case for Approach 1, though there it
appears that the sum over $l$ of $2\Le +1$ always increases.

{\it Conclusion.---} There is a substantial literature on
  the harmonic approach to PTAs (see~\cite{allen2024harmonic} and
  references therein).  These represent the mean correlations between
  pulsars as a sum of Legendre polynomials of the cosine of the
  pulsar-pair separation angle.  The coefficients in that expansion characterize
  the correlations.

  Here, we have shown two ways to estimate these coefficients from
  measured redshift or timing-residual measurements. The first
  approach minimizes the expected variance away from the mean, for a
  single coefficient ``in isolation''.  The second approach minimizes
  the square of the mean deviation away from the expected value, where
  the measure on the quadratic form is constructed in the same way as
  for a conventional $\chi^2$ statistic.  Of greatest interest to us
  is not the estimators themselves but rather their variances.

  In both cases, the variance is written as a ratio, where the
  numerator $\langle c_l \rangle^2$ is the square of the expected
  value, and the denominator is an effective number of degrees of
  freedom.  This denominator, in turn, is the product of an effective
  number of angular degrees of freedom and an effective number of
  frequency degrees of freedom.  Both of these depend upon $l$, but
  the effective number of angular degrees of freedom is determined
  entirely by the geometry of the PTA, whereas the effective number of
  frequency bins depends upon the spectral properties of the GWB and
  pulsar noise.  The corresponding formulas for first approach are
  given in \eq{e:var_c2}, \eq{e:Le1} and \eq{e:NF1}, and for the
  second approach in \eq{e:var_c4} \eq{e:Le2} and \eq{e:NF2}.

  The ``harmonic-space'' analysis done here closely parallels a
  corresponding ``position-space'' analysis done in~\cite{OptimalHD},
  and shares many similar features.  Both construct frequency-weighted
  estimators of the HD correlation.  In comparison with the ``white,
  zero-lag'' estimators of the HD correlation used
  in~\cite{AllenHDVariance,AllenRomanoPTA}, these permit a further
  reduction in the variance.  Indeed, by increasing the effective
  number of signal-dominated frequency bins, the effects of cosmic
  variance can in principle be reduced as much as desired.  Such an
  increase can be achieved either by adding additional ``quiet''
  pulsars to the PTA, or alternatively, by increasing the observation
  time.  Similar conclusions have been reached
  in~\cite{pitrou-cusin:2024}.

  We believe that applying these approaches to PTA posteriors will
  provide a useful tool to characterize and quantify the HD
  correlations and their deviations from the HD curve.

\bibliography{references}
\end{document}